\documentclass[11pt,english]{article}
\pdfoutput=1
\usepackage{jheppub}
\usepackage{amsfonts}
\usepackage{bbm}
\usepackage{babel}
\usepackage{verbatim}
\usepackage{mathrsfs}
\usepackage{amsmath,amssymb,mathrsfs}
\usepackage{hyperref}
\everymath{\displaystyle}

\renewcommand{\d}{\delta}
\newcommand{\bea}{\begin{eqnarray}}
\newcommand{\eea}{\end{eqnarray}}
\newcommand{\bel}{\begin{align}}
\newcommand{\eel}{\end{align}}
\newcommand{\p}{\partial}
\newcommand{\bes}{\begin{subequations}}
\newcommand{\ees}{\end{subequations}}
\newcommand{\non}{\nonumber}

\title{Uniqueness of Galilean Conformal Electrodynamics and its Dynamical Structure}

\author[a]{Kinjal Banerjee,} \author[a]{Rudranil Basu,} \author[a]{Akhila Mohan}
\affiliation[a]{BITS-Pilani, KK Birla Goa Campus, NH 17B, Bypass Road, Zuarinagar, Goa, India 403726}

\emailAdd{kinjalb@gmail.com, rudranilb@goa.bits-pilani.ac.in, mohan.akhila90@gmail.com}

\abstract{
We investigate the existence of action for both the electric and magnetic sectors of Galilean Electrodynamics using Helmholtz conditions. We prove the existence of unique action in magnetic limit with the addition of a scalar field in the system. The check also implies the non existence of action in the electric sector of Galilean electrodynamics. Dirac constraint analysis of the theory reveals that there are no local degrees of freedom in the system. Further, the theory enjoys a reduced but an infinite dimensional subalgebra of Galilean conformal symmetry algebra as global symmetries. The full Galilean conformal algebra however is realized as canonical symmetries on the phase space. The corresponding algebra of Hamilton functions acquire a state dependent central charge.}

\begin{document}

\maketitle

\section{Introduction}
The role that Conformal Field Theories (CFT) play in the modern understanding of Quantum Field Theories (QFT) in general can never be overemphasized. This connection basically stems from the Wilsonian point of view of looking at all QFTs as some deformations of some CFTs. Therefore, the highly ambitious goal of scanning all QFTs (renormalizable or not) boils down to finding equivalence classes containing CFTs from which they flow from or flow to depending on whether the deformations are irrelevant or relevant.

Most tractable of CFTs are those in 2 space-time dimensions. This is mainly because they have the infinite dimensional Virasoro algebra as conformal symmetry generators. Owing to this, 2D CFTs are integrable. This means that a huge number of analytical predictions can be made for these systems, not exclusively limited to multi-point correlators. And much of this comes about without a detailed knowledge of the specific dynamics of the particular model.

In higher dimensions however, the conformal symmetry groups are finite dimensional. There are however only a few higher dimensional superconformal field theories with infinite hidden symmetries, thereby making them extremely important in making nonperturbative predictions. For example the 4 dimensional $\mathcal{N}=4$ super Yang-Mills theory or the 3 dimensional $ \mathcal{N} =6$ Chern Simons thoery (ABJM) enjoy the infinite dimensional Yangian symmetry (see eg. \cite{Beisert:2010jr} for a quantum treatment and \cite{Beisert:2018zxs} for a more recent classical one). It is natural therefore to search for newer higher dimensional CFTs with large number of global symmetries. Interestingly here, as with many physics problems, ideas from geometry provide us with some clues. It is known that Newton-Cartan (NC) manifolds \cite{Kuenzle:1972zw} of any dimensions, unlike Riemannian ones are allowed to have infinite number of `conformal isometries' \cite{Bagchi:2009my} \footnote{In the next section of this article we will make clear about what is meant by isometry for a non-Riemannian manifold}. This information itself is a natural invitation to construct QFTs on NC geometries. Evidently, such QFTs won't be compatible with the principles of special or general relativity. The natural question then is to ask whether they might be consistent with Galilean relativity. The answer is in the affirmative, for certain NC geometries. Similar results have also been reported in \cite{Banerjee:2014pya}. 

Galilean relativity is well known for centuries and is suited to describe dynamics at speeds much less than that of light. Galilean relativity is all about the following set of \textit{finite number} of symmetry generators: spatial and temporal translation, homogeneous spatial rotations and the transformation to another inertial frame (Galilean boost). However, when we allow the scale transformations:
$$ t \rightarrow \lambda t \, ,~~ ~~ x^i \rightarrow \lambda x^i$$ and associated special conformal symmetry (tailored for compatibility with Galilean relativity), a plethora of new and arbitrarily time dependent symmetry generators emerge. These, while still being consistent with Galilean relativity, give the Lie algebra of symmetry generators an infinite lift and match exactly with a class of isometries of NC manifolds.

Interestingly, there exists many different classes non-relativistic (non Lorentz invariant) kinematics other than the Galilean conformal one. 
For example, non relativistic QCD has been explored in \cite{Brambilla:2003nt}. 
One obvious well known departure from conventional conformal scaling is of Lifshitz type, that provide a suitable kinematical platform for a number of condensed matter systems \cite{Geracie:2014nka} (and references therein) ranging from quantum Hall systems to cold atoms:
$$ t \rightarrow \lambda^z t \, ,~~ ~~ x^i \rightarrow \lambda x^i .$$
Here $z$, the Lifshitz dynamical exponent. Lifshitz scalings have also been explored in the context of supersymmetric theories (\cite{Mohan:2019ohc},\cite{Chapman:2015wha},\cite{Gomes:2015cia} )
The $z=2$ case gives rise to Schrodinger symmetry group (the group of transformations that keeps the Schroedinger equation invariant).

Coming back to Galilean conformal symmetries, Galilean limit of electrodynamics in 4 dimensions has been being studied for a long time \cite{levy} \footnote{A striking feature observed back then is that at the Galilean limit, the (sourced) equations of motion of electrodynamics fall into two diconnected sectors which cannot be mapped from each other by any continuous set of Gailean transformations. These sectors are named as the electric and the magnetic sectors}. But, it is only recently that field theories (electrodynamics being the first non-trivial one \cite{Duval:2014uoa}) with Galilean conformal symmetry have gained much attention. This is mainly fueled by the existence of infinite number of Galilean conformal global symmetry generators in spacetime dimensions $>2$, as part of NC isometries \cite{Duval:2009vt}, \cite{Bagchi:2009my}. In \cite{Bagchi:2014ysa} it was shown, at the level of equations of motions, that electrodynamics at the non-relativistic (Galilean) approximation, indeed is covariant under the infinite number of Galilean conformal transformations. This approach was subsequently carried out for interacting gauge theories (pure Yang-Mills and other matter coupled gauge theories) \cite{Bagchi:2015qcw}, \cite{Bagchi:2017yvj}.

While the large body of studies on Galilean field theories mentioned above are at the level of equations of motion, a need for action formalism seems to be important from the perspective of constructing a quantum theory. However in a few cases like \cite{Banerjee:2015rca} such analysis has been carried out at the level 
of action. This is more so for interacting theories. In \cite{Bergshoeff:2015sic}, such an action for the magnetic sector was first presented, with more anaysis of dynamical aspects presented in \cite{Festuccia:2016caf}. Notably in these works, a new dynamical field in the system was introduced over and above the field content of the magnetic sector of Galilean electrodynamics. Moreover in \cite{Bergshoeff:2015sic} and other following works, the technical tools were developed for coupling electrodynamics (or any other fields or strings \cite{Batlle:2016iel}, \cite{Kluson:2018egd}) to generic NC background \cite{Kluson:2018zlo}, \cite{Festuccia:2016awg}. These ideas were used for implementing gravitational theories Galilean or NC manifolds \cite{Bergshoeff:2016xsr}. 

One of the main goals of the present paper is to systematize the Galilean field theory equations of motion so as to connect with the action principle, ie. to probe the feasibility of an inverse variational problem for these system  of equations. This problem in general has been studied for a long time (see \cite{Morandi:1990su} and references therein) for ordinary differential equations (ie. system of particles)\cite{Davis : 1928, kinjal:2016} as well as for  partial differential equations (ie. field theories) \cite{Henneaux:1984ke}. The necessary and sufficient conditions for achieving this are given by a set of conditions referred to as Helmholtz conditions. Here we propose an algorithm for finding an action functional (if it exists) which give us the required equations of motion. Briefly, for an action to exist, the equations of motion should satisfy the Helmholtz conditions and obey the underlying Galilean symmetries.

The other novel feature of the present project is regarding the dynamical realization of the infinite dimensional Galilean conformal symmetry. This is basically motivated by the question whether Galilean conformal symmetries are respected at the quantum level via Ward identities. Ward identities address most straightforwardly, the question of validity of classical symmetries at the quantum level and any breaking of those appear as anomalies thereof. To this end we make a few observations at the classical level. The first one is that, while going to the action formulation of Galilean electrodynamics by introducing new degrees of freedom (as in \cite{Bergshoeff:2015sic}), a huge symmetry reduction happens. The infinite dimensional Abelian ideal, along with the global finite dimensional Galilean conformal symmetries do survive as symmetries of action. Secondly, via Dirac constraint analysis, we see that the new Galilean conformal theory is devoid of any propagating degrees of freedom. Interestingly though, all the symmetries of equations of motion (before introduction of a new field) now remain alive as Hamilton vector fields on the phase space. And algebra of the corresponding Hamilton functions on the phase, acquire a state dependent central charge.

The paper is organized as follows. In Section 2, we review the basics of Galilean Electrodynamics including the Galilean conformal symmetry and its action on the fields. We also write down the equations of motion in two sectors of the electrodynamics called electric and magnetic sectors. In Section 3 we apply Helmholtz conditions to the equations of motion to check for the existence of action in both electric and magnetic sectors. We find that an unique action exists in magnetic limit with the addition of an extra scalar field. However no such action exists for the electric sector even with the addition of an extra scalar or an extra vector field.  In the last section we study the dynamical structure of Galilean electrodynamics using the action obtained in the magnetic limit. After obtaining the Hamiltonian we carry out the constraint analysis of the theory via the Dirac algorithm. We show that there are no local degrees of freedom present and discuss the fate of the original infinite set of symmetries in the phase space of our theory. In the appendix B, we attempt to construct an action for Carrollian electrodynamics using the procedure described in Section 2.

\section{Galilean Electrodynamics}
Galelian electrodynamics is the answer to the question whether there exists a consistent and physically meaningful non-relativistic theory of classical electromagnetism \cite{Bellac:1973}.
However, instead of a single non relativistic limit, there are two different limits
called electric and magnetic limits depending on the dominance of electric and magnetic effects respectively. Here, magnetic and electric limits are expressed in scaling of space-time coordinates and similarly the difference in scaling of the gauge field components.


\subsection{Galilean conformal symmetry}
In $d+1 (d>1)$ dimensional Minkowski space-time, conformal symmetry generators are the space-time translations ($\tilde{P}_i, \tilde{H} $), rotations ($\tilde{J}_{ij}$) and boosts ($\tilde{B}_i$) as well as Dilatation  ($\tilde{D}$) and special conformal transformations ($\tilde{K}_\mu$). As vector fields, these are: 
\begin{eqnarray} \label{mink_gen}
 && \tilde{P}_i = \partial_i ~ , ~ \tilde{H}=-\partial_t ~, ~ \tilde{J}_{ij} = x_i \partial_j - x_j \partial_i\nonumber \\
&&  \tilde{B}_i = x_i \partial_t + t \partial_i  ~ , ~ \tilde{D}= -(t.\partial_t + x^i\p_i) \nonumber\\ 
&&  \tilde{K} = -2x_\mu(x.\partial)+ (x.x)\partial_\mu.
\end{eqnarray}
Going to the Galilean limit, means scaling space and time coordinates inhomogeneously, taking the non-relativitic limit in the scaling parameter:
\begin{eqnarray} \label{xtlim}
 x_i \rightarrow \epsilon x_i  , t \rightarrow t ,   \epsilon \rightarrow 0 \label{scaling}
\end{eqnarray}
 and regularizing the generators.
 
Understandably the Galilean generators thus found by these scaling limits are different than the relativistic ones. The explicit forms of these are:
\begin{eqnarray}\nonumber \label{fgca_gen}
&& {P}_i = \partial_i  ~ ~ ~; ~ ~ ~ {H}=-\partial_t \\ \nonumber
&& {J}_{ij} = - x_i \partial_j + x_j \partial_i   ~ ~ ~; ~ ~ ~ {B}_i =t \partial_i\\ 
&& {D}= -x.\partial ~ ~ ~; ~ ~ ~ K = -(t^2 \partial_t + 2 x_i t \partial_i) ~ ~ ~; ~ ~ ~  {K}_i= t^2 \partial_i
\end{eqnarray}
We can organize some (all, apart from that of rotation) of the the generators in a more suggestive form as
\begin{subequations} \label{vfs}
\begin{eqnarray} 
 L^{(n)} &=& -t^{n+1} \partial_t - (n+1)t^n x_i \partial_i  \hspace{2cm} (n= -1,0,1 \implies H,D,K) \\
M_i^{(n)} &=& t^{n+1} \partial_i  \hspace{4.5cm} (n=-1,0,1 \implies P_i , B_i , K_i).
 \end{eqnarray}
 \end{subequations}
The Lie-brackets of these generators are given by:
\bea \label{gca1}
&& [L^{(n)}, L^{(m)}] = (n-m) L^{(n+m)}, \, [L^{(n)}, M^{(m)}_i] = (n-m) M_i^{(n+m)},\, [M^{(n)}_i, M^{(m)}_j]=0 \non \\
&& [L^{(n)}, J_{ij}] = 0, \quad [J_{ij}, M^{(n)}_k] = M^{(n)}_{[j} \delta^{}_{i]k}.
\eea
We name this $(d+2)(d+3)/2$ dimensional Lie algebra as finite Galilean conformal algebra (f-GCA). Note that, as expected, the $L$'s and the $M_i$'s transform in the scalar and fundamental vector representation of the `rotation' group $SO(d)$ generated by the $J_{ij}$. This, as a Lie algebra is distinct from relativistic conformal algebra and can alternatively be arrived upon by making an Inonu-Wigner contraction on the relativistic one.

Probably the most interesting aspect of Galilean physics comes about through the fact that, the above algebraic structure \eqref{gca1} closes for all integer values of $n$ and $m$; not just for the values $0, \pm 1$. This embeds the f-GCA into an infinite dimensional one. We will be working with this infinite dimensional version and will call it simply GCA. A couple of points of interest regarding GCA is that it has the circle deformation algebra as a subalgebra ($L^{(n)}$) and the arbitrary time dependent spatial translation generators ($M^{(n)}_i$) form an Abelian ideal. As a consequence of the later point this as a Lie algebra becomes non semi-simple, although it's parent relativistic conformal algebra is semi-simple.

With the infinite lift, a more compact way of writing \eqref{gca1} is by making the replacements in 
\bea \label{mlvf}
M_i^{(n)} \rightarrow M_{\xi} = \xi^i(t) \p_i &,& L^{(n)} \rightarrow L_{f} = -f(t)\p_t  - \dot{f}(t) x^i \p_i \non \\
 J_{ij} \rightarrow J_{\omega} &=& - \omega^{ij} (x_i \p_j - x_j \p_i)
\eea
 for a $SO(d)$ vector $\xi^i(t)$, a scalar $f(t)$, both of which are Laurent polynomials of $t$, and a constant antisymmetric matrix $\omega^{ij}$:
\bea \label{gcafunc}
[L_{f_1}, L_{f_2}] =  L_{\dot{f}_1 f_2 - f_1 \dot{f}_2} ~ , ~ [L_{f}, M_{\xi}] = M_{\dot{f}\xi - f \dot{\xi}} \non \\
\left[J_{\omega}, M_{\xi}\right] = M_{\zeta}  ~~ \mbox{where } \, \zeta^i = \omega^i{}_j \xi^j.
\eea
There is yet another, geometric way \cite{Duval:2009vt} of understanding Galilean conformal generators. As a first step towards this, we need to the go to formulation of non-relativistic space-time as non-Riemann space-time, namely a Newton-Cartan (NC) manifold $M$. An NC manifold is equipped with a degenerate, rank-2, contravariant, symmetric tensor $\gamma ^{\mu \nu}$ and a closed 
1-form $\theta_{\mu}$ which is in the Kernel of $\gamma^{\mu \nu}$. 

In analogy to conformal isometries of Riemann manifolds, we can define conformal NC `isometries' to be generated by vector fields $X$ such that for an arbitrary function $\psi$:
\bea \label{lie}
\pounds_X \gamma = \psi \gamma ~ , ~ \pounds_X \theta = -\frac{1}{2} \psi \theta.
\eea
The second condition above restricts $\psi$ as:  
\bea \label{gtheta}
d \psi \wedge \theta = 0.
\eea
At this point, let us take as the simplest example of $M$ as an open subset of $\mathbb{R} \times \mathbb{R}^d$, to be named as flat NC space-time from now on. Also we will work in a coordinate chart $(t, x^1, \dots , x^d)$ such that,
\bea
\gamma = \delta^{ij} \p_i \otimes \p_j ~,~ \theta = dt .
\eea
In this coordinate chart, \eqref{gtheta} implies that $\psi$ is a function of $t$ only. This also facilitates solving for $X$ as:
\bea \label{Xsolve}
X = f(t) \p_t + \left( \omega^i{}_j (t) x^j + x^i \dot{f}(t) + \xi^i(t) \right) \p_i ,
\eea
where $\omega_{ij}$ is anti-symmetric. Since $f, \xi^i$ and $\omega_{ij}$ are arbitrary functions of $t$, the algebra of conformal isometries of flat NC is infinite dimensional for any space-time dimensions. This is striking in a sense that conformal isometries in $3$ or higher dimensional Riemann manifolds are finite. Comparison of \eqref{mlvf} with \eqref{Xsolve} reveals that GCA generators actually consist of a subset of flat NC `isometries', while the former is projected onto the later by the restriction $\p_t \omega_{ij}(t) =0$.

\subsection{Action on fields} 
One of the main Kinematical ingredients required for the later part of the paper is the action of GCA generators on physical fields. For tensor fields these would be given by transformation rules under the above mentioned Galilean conformal diffeomorphisms. However we would not be directly working with tensor fields on flat NC manifolds, but rather with those transforming in fundamental representation of the `rotation' Lie-algebra $\mathfrak{so}(d)$ and simultaneously carrying definite scaling dimension (corresponding to the action of $D \equiv L^{(0)}$). This boils down to finding the scale-spin representation of the Lie-algebra \eqref{gca1}.

We define primary fields in a way analogous to relativistic conformal field theory. For example, for the case of $d=3$, our primary field $\Phi(t, x^i)$ at space-time origin $(0,0) \in \mathbb{R}\times \mathbb{R}^d$ should transform as:
\bea
\delta_{J^2} \Phi (0,0)= l(l+1) \Phi (0,0) ~,~ \delta_{D} \Phi (0,0)= \Delta\Phi (0,0).
\eea
$l, \Delta$ are the corresponding spin integer and scaling dimension respectively. Supplementary conditions of spatio-temporal transformation are implemented as:
\bea
\delta_H \Phi(t, x^i) = \p_t \Phi(t, x^i) ~, ~  \delta_{P_i} \Phi(t, x^i) = \p_i \Phi(t, x^i).
\eea
Highest weight representation is then defined by the primary $\Phi(0,0)$:
\bea
\delta_{L^{(n)}} \Phi(0,0) = 0 = \delta_{M^{(n)}_i} \Phi(0,0), \, ~ \forall n>0.
\eea
This condition is important for putting a lower bound for scaling dimensions of field content in a particular theory. Note that the fields found by acting on with the negative $L$ and $M_i$ modes will be in the usual CFT sense be named descendants.

The only part that remains in completely fixing the scale-spin representation is the way the Galilean boost generator $B_i = M^{(0)}_i$ acts of on the above defined primary field. As has been explored earlier \cite{Bagchi:2014ysa}, it is intimately tied with $\mathfrak{so}(d)$ spin multiplet that we consider in our theory. For the purpose of this paper, we consider the simple multiplet consisting only of $\mathfrak{so}(d)$ scalars and vectors. In this case, the action of Galilean boost is specified modulo a couple of c-numbers $a,b$:
\begin{align} \label{boostlaw}
\d_{B_i} \Phi^{\mathrm{scalar}}(0,0) = a \Phi^{\mathrm{vector}}_i(0,0) \, , \, \, \d_{B_i} \Phi^{\mathrm{vector}}_j(0,0) = b \delta_{ij}\, \Phi^{\mathrm{scalar}}(0,0) 
\end{align}
Now onward the explicit vector and scalar nature of the fields won't be specified. It will be evident from the index structure. The exact values of the undetermined constants $a,b$ appearing in \eqref{boostlaw} can be found either by (i)comparing with the Galilean limits of corresponding relativistic transformation laws or (ii) by imposing dynamical input through the explicit field theory we are concerned with. Here for the sake of brevity, we will focus on the later path. 

With the stage all set, one can now transport the transformation rules mentioned above from $(0,0) \in \mathbb{R}\times \mathbb{R}^d$ to any arbitrary point using the finite space-time translation generators to arrive at the following:

\begin{eqnarray}
\delta_{L^{(n)}} \Phi(t,x) &=& t^n (t\partial_t + (n+1)x^j \partial_j + (n+1)\Delta) \Phi(t,x)\nonumber \\
                    &-& b n(n+1) t^{n-1} x^i B_i(t,x)\nonumber \\
\delta_{L^{(n)}}\Phi_i(t,x) &=& t^n ( t\partial_t + (n+1)x^j \partial_j + (n+1)\Delta ) \Phi_i(t,x)\\ \nonumber
                    &-& a n (n+1) t^{n-1} x_i \Phi (t,x)\\ \nonumber
\delta_{M_i^{(n)}}\Phi{(t,x)} &=& -t^{(n+1)} \partial_i \Phi(t,x) + b (n+1) t^n \Phi_i(t,x) \\ \nonumber
\delta_{M_i^{(n)}} \Phi_j {(t,x)} &=& -t^{(n+1)} \partial_i \Phi_j(t,x) + a (n+1) t^n \delta_{ij} \Phi(t,x) \label{galconfinv} 
\end{eqnarray}
\subsection{Equations of motion}
A Galilean version of electrodynamics, intrinsically should be a theory of a vector field on an NC, or more specifically on a flat-NC manifold. The basic assumption behind this is the existence of the potential formulation. The potential formulation is plausible because the Maxwell's equations $dF = 0$, ie the field strength 2-form being closed is independent of any other structure than differentiability. For almost all practical topologies, it is therefore imminent that a 1-form $A$ exists. However the equation $\star d \star dA  \equiv J$ involving sources is valid on a Riemann manifold; and does not have a straightforward parallel in the Galilean framework or more generally on an NC manifold. In order to ease this tension, resorting \cite{Bagchi:2014ysa} to the point of view of going to Galilean regime from the relativistic (Minkowski) regime via the limit prescription \eqref{xtlim} proves to be fruitful. 

As a first step, one splits the potential 1-form $A_{\mu}$ into temporal and spatial parts, $A_t, A_i$ and scales them in accordance with \eqref{xtlim}. However, depending upon the causal nature of the Minkowski field $A_{\mu}$, couple of such limits are there \cite{levy}:
\begin{eqnarray}
 \text{Electric limit} : A_t \rightarrow A_t ; A_i \rightarrow \epsilon A_i \\ \label{electric1}
 \text{Magnetic limit} : A_t \rightarrow \epsilon A_t ; A_i \rightarrow A_i \label{magnetic1}
\end{eqnarray}
We can see from the above equations and \eqref{xtlim} that when $\epsilon\rightarrow0$, electric effects suppress the magnetic ones in the limit \eqref{electric1} and vice versa in the limit \eqref{magnetic1}. Hence two sectors of Galilean electrodynamics emerge, named respectively the `electric sector' and the `magnetic sector' according to the limit prescriptions giving rise to them.

Starting from the parent relativistic equation $ \Box A_{\mu} - \p_{\mu} \p \cdot A = 0$ (turning off the source for now), the explicit equations of the above mentioned sectors descend from the relativistic equation as:
\begin{eqnarray}
 \text{Electric sector  }: T_0 &:=&\partial^i \partial_i A_t =0 \nonumber \\
			T_i &:=& \partial^j \partial_j A_i - \partial_i \partial_j A^j - 
				\partial_t \partial_i A_t = 0 \label{eomelectric}\\
 \text{Magnetic limit  }: T_0 &:=&\partial_i \partial_t A^i +(\partial^j \partial_j)A_t =0 \nonumber \\
		      T_i &:=& \partial^j \partial_j A_i - \partial_i \partial_j A^j = 0 \label{eommagnetic} 
\end{eqnarray}
where the labels $T_0$ and $T_i$ refer to equations of motion for $A_t$ and $A_i$ respectively in
both cases. It was proved in \cite{Bagchi:2014ysa} that both the Electric and the Magnetic sector equations of motion enjoy Galilean conformal symmetry under the transformation rules spelled out in \eqref{galconfinv} for spatial dimension $d =3$ and common conformal dimension $\Delta =1$.

\section{Towards an action formulation}\label{helmholtzaction}
\subsection{The Helmholtz Conditions}

Most of the theories in classical physics start from an action ($S$) and a Lagrangian associated with it. Then the equations of motion are derived via variational principle and the Hamiltonian obtained from the Lagrangian via Legendre transformations. Sometimes, however, we may have a situation (like in Galilean and Carrollian electrodynamics)  where we know the equations of motion but not the Lagrangian from which they have been obtained. The question that naturally arises in such cases is that: given a set of equations of motion, is it possible to find a Lagrangian corresponding to them? In other words we would like to know  whether a given set of second order partial differential equations governing the dynamics of a physical system can be obtained as Euler-Lagrange equations of some Lagrangian function. The necessary and sufficient conditions for this to be so are known as Helmholtz conditions. This ``inverse'' problem of calculus of variations has been studied in mathematics literature 
\cite{davis2,jessedouglas}.

Consider a theory described in terms of fields $u^B$ (here the indices $A,B, \dots$ indicates the number of fields and can go from
$1$ to $N$) whose equations of motion are denoted by $T_A$. The necessary and sufficient conditions for an action
functional $S[u^B] = \int d^n x \mathcal{L}(u^B,u_a^C,u^D_{ab},x^a)$ corresponding to these equations of motion to
exist are given by the Helmholtz conditions \cite{Henneaux:1984ke}
\begin{eqnarray}
 \frac{\partial T_A}{\partial(u_{ab})^B} &=& \frac{\partial T_B}{\partial (u_{ab})^A} \label{helmholtz1}\\
 \frac{\partial T_A}{\partial {u_a}^B} + \frac{\partial T_B}{\partial {u_a}^A} &=& 2 \partial_b \frac{\partial
T_B}{\partial ( u_{ba})^A } \label{helmholtz2} \\
 \frac{\partial{T_A}}{\partial {u^B}} &=& \frac{\partial T_B}{\partial u^A} - \partial_a \frac{\partial T_B}{\partial
{u_a}^A} + \partial_a \partial_b {\frac{\partial T_B}{\partial (u_{ab})^A}} \label{helmholtz3}
\end{eqnarray}
where $u_a^A$and $u_{ab}^A$ are the first and second derivatives of $u^A$. 

In the context of Galilean electrodynamics, where we have the equations of motion, it is natural to see if they come from an action principle. In order to do that systematically, we will implement the following steps.
\begin{enumerate}
	\item Firstly we will pass the equations of motion through Helmholtz criteria. If they satisfy the criteria, we go down to step \ref{gicheck} below. If these are not satisfied, then we go to step \ref{step2}.
\item \label{step2} Append the system of equations minimally with new fields with well defined GCA transformation rules, such that the equations still remain linear and now satisfy the Helmholtz criteria.
\item The set of equations thus found will then be further constrained by requiring them to give back the original Galilean electrodynamics equations \eqref{electric1},\eqref{magnetic1} when the newly introduced field(s) is (are) set to zero.
\item \label{gicheck} Finally these should possess (at least the f-GCA \eqref{vfs} part of) the Galilean conformal symmetry. For this, we proceed by checking whether the equations still continue to hold with infinitesimally transformed field variables $\Phi(t,x)$. Stated differently, if an equation of motion has the symbolic form: 
$ \square \Phi(t,x) =J$  (for non-dynamical $J$) and if :
\bea
\square \delta_{ \bigstar} \Phi(t,x) =0 ~~~ \mbox{on -shell},
\eea
then the equation of motion is said to be invariant under the symmetry generated by $\bigstar \in \mathrm{GCA}$. Geometrically speaking, the symmetry transformations $\delta_{ \bigstar} \Phi(t,x)$ is a tangent vector on the space of solutions.
\end{enumerate}

The explicit program of investigating for an action in both the magnetic and electric sectors of Galilean electrodynamics is discussed below while we touch upon the topic of Carrollian conformal electrodynamics in Appendix \ref{carrollian}.   

\subsection{Action for Galilean Electrodynamics}

The equations of motion for the magnetic and the electric limit of Galilean electrodynamics 
are given by \eqref{eommagnetic} and \eqref{eomelectric} respectively. The action for the magnetic case was obtained  in \cite{Bergshoeff:2015sic} while no action for the electric limit has been obtained so far. In this section, we will develop a systematic way of obtaining the action or of checking whether an action exists. 

Consider the magnetic limit equations of motion \eqref{eommagnetic}. As mentioned before, $T_0$ refers to the equation of motion for $A$ and $T_i$ to the  equations of motion for $A_i$. From the Helmholtz conditions given above it can be clearly seen that no Lagrangian can exist whose Euler-Lagrange's equations of motion are given by \eqref{eommagnetic} since
\begin{eqnarray}
&& \frac{\partial T_0}{\partial (A_i)_{ab}} = -1  ~ ~ ~ \mbox{for } a=i \mbox{ and } b=t \nonumber \\
\mbox{while} ~ ~ && \frac{\partial T_i}{\partial (A)_{ab}} = 0 ~ ~ ~ \mbox{for all } a,b \nonumber
\end{eqnarray}
Hence the first Helmholtz condition \eqref{helmholtz1} is violated. 

Similarly consider the electric limit equations of motion \eqref{eomelectric}.
\begin{eqnarray}
&& \frac{\partial T_0}{\partial ({A_i})_{ab}} = 0 ~ ~ ~ \mbox{for all } a,b \nonumber\\
&& \frac{\partial T_i}{\partial ({A})_{ab}} =-1 ~ ~ ~ \mbox{for } a=t \mbox{ and } b=i. \nonumber
\end{eqnarray}
Again the first Helmholtz condition is violated and we cannot obtain a Lagrangian whose equations of motion will correspond to the equations of motion of electric limit.

As per the strategy mentioned above, let us modify the equations of motion minimally by adding a  scalar field $B$. The most general set of equations of motion for $A, A_i, B$  which are linear in the fields and quadratic in derivatives are given respectively by
\begin{eqnarray}
T_0 &=& a_1 \partial_j \partial_j A + b_1 \partial_j \partial_t A_j + a_2  \partial_t \partial_t A +
 c_1 \partial_j \partial_j B + c_2 \partial_t \partial_t B = 0 \nonumber\\
T_i &=& b_2 \partial_j \partial_j A_i + b_3 \partial_i \partial_j A_j + b_4   \partial_t \partial_t A_i + a_3 \partial_t
\partial_i A + c_3 \partial_i  \partial_t B = 0  \nonumber \\ 
T_{B} &=& c_4 \partial_j \partial_j B + c_5 \partial_t \partial_t B + a_4  \partial_t \partial_t A + 
a_5 \partial_j \partial_j A + b_5  \partial_j \partial_t A_j = 0 \label{eommostgeneral}
\end{eqnarray}

We have to check whether there exists values for the undetermined constants $a_1, b_1, c_1$ etc which are consistent with the Helmholtz conditions and Galilean conformal invariance and which gives back the equations \eqref{eomelectric} or \eqref{eommagnetic} when $B$ is set to zero. It can be easily verified that:
\begin{eqnarray}
\frac{\partial T_0}{\partial (B)_{ab}}&=& c_1 ~ ~ \mbox{for} ~ ~ a=b=j \\ \nonumber
                                         &=& c_2  ~ ~ \mbox{for} ~ ~  a=b=t \\\nonumber
\frac{\partial T_{B}}{\partial (A)_{ab}}  &=& a_4  ~ ~ \mbox{for} ~ ~  a=b=t \\ \nonumber
                                                &=& a_5  ~ ~ \mbox{for} ~ ~  a=b=j
\end{eqnarray}
Since the first Helmholtz condition demands $\frac{\partial T_0}{\partial (B)_{ab}} = \frac{\partial T_{B}}{\partial (A)_{ab}} $ we obtain the relations $c_1 = a_5$ and $c_2 = a_4$. 

Similarly, from $\frac{\partial T_0}{\partial (A_i)_{ab}} = \frac{\partial T_i}{\partial {(A)}_{ab}} $ we get $b_1 = a_3$ and from $\frac{\partial T_i}{\partial (B)_{ab}} = \frac{\partial T_B}{\partial {(A_i)}_{ab}} $ we have $c_3 = b_5$.  

Using the above constraints, the equations of motion can be modified as,
\begin{eqnarray}
T_0 &=& a_1 \partial_j \partial_j A + a_3 \partial_j \partial_t A_j + a_2 \partial_t \partial_t A + a_5 \partial_j \partial_j B +
 a_4 \partial_t \partial_t B = 0 \nonumber\\
T_i &=& b_2 \partial_j \partial_j A_i + b_3 \partial_i \partial_j A_j + b_4  \partial_t \partial_t A_i + 
a_3 \partial_t \partial_i A + b_5 \partial_i \partial_t B = 0 \nonumber\\
T_{B} &=& c_4 \partial_j \partial_j B + c_5 \partial_t \partial_t B + a_4 \partial_t \partial_t A + 
a_5 \partial_j \partial_j A + b_5 \partial_j \partial_t A_j = 0 \label{eomgeneral}
\end{eqnarray}

At this stage these remaining parameters can be chosen arbitrarily. We will fix these parameters by demanding that these equations give back the magnetic(electric) limit equations of motion once the field $B$ is set to zero. This sets up us with the following constraints:
\bea \label{ab_coeffs}
b_2= - b_3 ~, ~ b_5 = -a_5 ~ , ~ a_1 = a_2 = a_3 = a_4 = b_4 = 0
\eea
with the remaining parameters appearing in \eqref{eomgeneral} are still arbitrary.

On the other hand, it can be easily seen that that we cannot obtain the electric limit of the equations of motion \eqref{eomelectric} from the above general equations \eqref{eomgeneral} by any choice of the parameters. This implies that we cannot construct an action for electric limit by enhancing the field content with an extra scalar field. We will come back to this case later.

After imposing the Helmholtz conditions the most general set of equations, which would give the magnetic limit once $B$ is set to zero, is given by
\begin{eqnarray}\nonumber \label{mageq}
T_0 &=& a_5 \partial_j \partial_j B  = 0\\
T_i &=& b_2 \partial_j \partial_j A_i -b_2 \partial_i \partial_j A_j  -a_5\partial_i \partial_t B = 0 \\ \nonumber
T_{B} &=&  c_4 \partial_j \partial_j B+ c_5 \partial_t \partial_t B +  
a_5\partial_j \partial_j A - a_5\partial_j \partial_t A_j   = 0
\end{eqnarray} 

The next part would be to check whether these equations are consistent with Galilean conformal invariance. 
For this we would need the transformation rules of the fields under the action of GCA generators. However, instead of just having a scalar and a vector as in the general prescription \eqref{galconfinv}, we here have two scalars ($A,B$) and a vector field $A_i$. Therefore adapting suitably the rules of \eqref{galconfinv} in the present scenario, we write below the transformation rules under GCA: 
\begin{eqnarray}\label{galconfinv2}
\d_{L^{(n)}}A(t,x)&=& t^n [t\partial_t + (n+1)x^j \partial_j + (n+1)\Delta ] A(t,x)\nonumber \\\nonumber
                    &-& b n(n+1) t^{n-1} x^i A_i(t,x)  \\ \nonumber
\d_{L^{(n)}} B(t,x)&=& t^n [t\partial_t + (n+1)x^j \partial_j + (n+1)\Delta] B(t,x)\\\nonumber
                    &-& b' n(n+1) t^{n-1} x^i A_i(t,x)\\
\d_{L^{(n)}}A_i(t,x) &=& t^n [ t\partial_t + (n+1)x^j \partial_j + (n+1)\Delta ] A_i(t,x)\\ \nonumber
                    &-& a n (n+1) t^{n-1} x_i A(t,x)+  a' n (n+1) t^{n-1} x_i B(t,x)\\\nonumber
\d_{M_i^{(n)}}A{(t,x)} &=& -t^{(n+1)} \partial_i A(t,x) + b (n+1) t^n A_i(t,x)  \\\nonumber
\d_{M_i^{(n)}}B{(t,x)} &=& -t^{(n+1)} \partial_i B(t,x) + b' (n+1) t^n A_i(t,x) \\\nonumber
\d_{M_i^{(n)}} A_j{(t,x)} &=& -t^{(n+1)} \partial_i A_j(t,x) + a (n+1) t^n \delta_{ij} A(t,x) - a' (n+1) t^n \delta_{ij} B(t,x)                     
\end{eqnarray}
where $a, b, a', b'$  are arbitrary constants. We now move on to the last part of program mentioned in the last section, ie, finding if \eqref{mageq} is invariant under \eqref{galconfinv2}.

It turns out that the set of equations \eqref{mageq}, which is an augmentation over the Galilean magnetic sector theory, is invariant under the action of $M_i^{(n)}, \forall n \in \mathbb{Z}$ and the action of $L^{(n)}, ~\mbox{ for } n=0, \pm 1$, if and only if the following conditions are met: 

\bea \label{parameters}
  a_5 =- a' b_2 \ &,& \  b'= 0 = a \non \\
  \Delta = d-2 \ &,& \   \Delta =1 \\
   c_5 =  a' a_5 \ &,& \  b = -1 \non
\eea
The condition, $\Delta = 1$ implies the invariance of the equations of motion only in $d=3$. This is consistent with the fact that relativistic Maxwell theory is classically conformal in 3 spatial dimensions, with the vector potential being a conformal dimension 1 field. Another feature to learn from \eqref{parameters} is that all the parameters ($a,b,b'$) except $a'$ that specify the representation of GCA (for given conformal weight and spin of the fields) are determined from the criteria of invariance.

Using the relations \eqref{parameters} in \eqref{mageq} we get a set of equations of motion which are invariant under the actions of $M^{(n)}_i$ and $L^{(-1)} , L^{(0)}, L^{(1)}$:

\begin{eqnarray} \label{magB}
T_0 &=& \partial_j \partial_j B = 0 \nonumber\\
T_i &=& \partial_j \partial_j A_i -  \partial_i \partial_j A_j 
+  \partial_i \partial_t B = 0 \\
T_B &=&  \partial_t \partial_t B +  \partial_j \partial_j A 
-  \partial_j \partial_t A_j = 0\nonumber. 
\end{eqnarray}
While writing the above we have made a trivial redefinition $B \rightarrow a' B$ (this effectively sets $a' = 1$ in \eqref{galconfinv2}) and \eqref{parameters}. 

The action which would give these equations of motion can be easily constructed and is give by
\begin{eqnarray}
S = \frac{1}{4} \int d^3 x \,dt \left[  \partial_t B \partial_t B + 2 \partial_j A \partial_j B - \frac{1}{2}F_{ij}F_{ij} - 2\partial_i B \partial_t A_i  \right]  \label{lagrangianmagnetic}
\end{eqnarray}
where $F_{ij} = \partial_i A_j - \partial_j A_i$. Also we would assume the fields to have appropriate fall-off conditions so that the vaiational principle works at spatial boundary. It can be easily seen that this is the same action proposed in \cite{Bergshoeff:2015sic} obtained from very different means. Hence it is proved that the theory that is a minimal extension to Galilean magnetic sector of electrodynamics has a unique action.

We conclude the discussion of the magnetic sector first by revisiting our earlier observation that the equations \eqref{eommagnetic} are invariant under full GCA. However the modified dynamics, which is consistent with an action principle, is invariant under a truncated but still infinite dimensional subalgebra of GCA. This subset of generators contain the Hamiltonian ($L^{(-1)}$), dilatation ($L^{(0)}$) and temporal part of special conformal transformation ($L^{(+1)}$) besides the momenta ($M^{(-1)}_i$), Galilean boost ($M^{(0)}_i$), spatial components of special conformal transformation ($M^{(+1)}_i$) and any arbitrary time dependent spatial translations ($\{M^{(n)}_i| i \in \mathbb{Z} \backslash \{0, \pm 1\} \}$).
\subsection{Nonexistence of Action for Electric Limit}

As discussed above, we cannot satisfy the Helmholtz conditions for electric limit even after adding an new scalar field. Hence no action formulation can exist for that case. Let us see whether an introducing a vector field $B_i$ allows us to construct an action. Following the same strategy detailed above, the most general equations of motion which are linear in the fields are
\begin{eqnarray}
 T_0  &=& a_1 \partial_j \partial_j A + a_2 \partial_t \partial_t A + b_1 \partial_j \partial_t A_j + 
          c_1 \partial_j \partial_t B_j = 0 \nonumber \\
T_{A_i} &=& a_3 \partial_i \partial_t A  + b_2 \partial_j \partial_j A_i  + b_3\partial_i \partial_j A_j 
             + b_4 \partial_t \partial_t A_i  + c_2 \partial_j \partial_j B_i
          + c_3 \partial_j \partial_i B_j + c_4 \partial_t \partial_t B_i = 0 \nonumber \\
T_{B_i} &=& a_4 \partial_i \partial_t A + b_5 \partial_t \partial_t A_i  + b_6 \partial_j \partial_j A_i  
             + b_7  \partial_i \partial_j A_j   + c_5 \partial_t \partial_t B_i  
         + c_6 \partial_j \partial_j B_i  + c_7 \partial_j \partial_i B_j =0 \nonumber\\ \label{eomgeneralelectric}
\end{eqnarray} 
The relations among these undetermined constants coming from the Helmholtz conditions are $b_1 = a_3$, $c_1 = a_4$, 
$c_2 = b_6$, $c_3 = b_7$ and $c_4 = b_5$ thereby reducing the equations \eqref{eomgeneralelectric} to
 \begin{eqnarray}
  T_0  &=& a_1 \partial_j \partial_j A + a_2 \partial_t \partial_t A + a_3 \partial_j \partial_t A_j 
           + a_4 \partial_j \partial_t B_j = 0 \nonumber\\
  T_{A_i} &=& a_3 \partial_i \partial_t A  + b_2 \partial_j \partial_j A_i  + b_3 \partial_i \partial_j A_j 
              + b_4 \partial_t \partial_t A_i  + b_6 \partial_j \partial_j B_i 
               + b_7 \partial_j \partial_i B_j + b_5 \partial_t \partial_t B_i = 0 \nonumber\\
  T_{B_i} &=& a_4 \partial_i \partial_t A + b_5 \partial_t \partial_t A_i  + b_6 \partial_j \partial_j A_i  
             + b_7  \partial_i \partial_j A_j   + c_5\partial_t \partial_t B_i 
             + c_6 \partial_j \partial_j B_i  + c_7 \partial_j \partial_i B_j =0 \nonumber\\ \label{eomelectric2}
\end{eqnarray}
To get the electric limit from these equations when $B_i$ is zero we need to have $b_6=-b_7=-a_4$, and $a_2,a_3,b_2,
b_3, b_4, b_5$ have to be zero, while the other parameters remain arbitrary. 
However, it can be checked using \eqref{galconfinv1} that $T_0$ is not invariant under $M_i^{(n)}$ and $L^{(n)}$. Here, $A$ will transform under the scalar transformation rule and $A_i$ and $B_i$ will transform under the vector field transformation rule \footnote{Since the field contents of the action are $ A$, $A_i$ and $B_i$ , the terms in proportional to the scalar field B vanishes in the transformations (or in this case, we can set $a'=0$ in the \eqref{galconfinv1})}. Hence it is not possible to construct an action which would give electric limit and be invariant under Galilean conformal transformations even after enhancing the set of field with a vector field.

In this section we have seen that Helmholtz conditions provide a very useful tool in constructing actions once equations of motion are known or to prove that actions will not exist. We shall follow the same strategy for the Carrollian limit for electrodynamics in the appendix.
\section{Dynamical structure}
We have seen the existence of a unique action for the magnetic sector with a necessary and sufficient introduction of a scalar, linearly in the system. While still Galilean conformal invariant, the new dynamics does not enjoy invariance under full GCA, but under a truncated yet infinite dimensional subalgebra of GCA. It is therefore imperative that we probe deeper into dynamical implications of these symmetries and look into corresponding Hamilton functions and conserved charges. As for the case of conserved charges, this study is important as being a precursor to the quantum theory, where we would be posing the questions like whether the Galilean conformal invariance is protected at the level of Ward identities. On the other hand we will encounter Hamiltonian functions (generators of canonical transformation on phase) which are not conserved in time.

To start off the section, we will carry out the constraint analysis of the theory following Dirac's algorithm, to identify the gauge invariance and redundant degrees of freedom. Following that, a covariant phase space \cite{Crnkovic:1986ex} analysis of symmetry generators will be carried out.

For the sake of the present section we restate the Lagrangian of magnetic sector Galilean electrodynamics \eqref{lagrangianmagnetic}:
\begin{align}
\label{Lmag}
L=\int d^d x \left[\frac{1}{4}\dot{B}^{2}+\frac{1}{2}\partial_{i}A.\partial_{i}B-\frac{1}{8}F_{ij}F^{ij}-\frac{1}{2}\dot A_{i}\partial_{i}B \right]
\end{align}
and the corresponding equations of motion: 
\bes \label{meom}
\begin{align}
\label{mag1}
& \partial_{t}\partial_{j}A_{j}-\partial_{j}^{2}A-\ddot B=0\\
\label{mag2}
& \partial_{i}\dot B-\partial_{i}\partial_{j}A_{j}+\partial_{j}^{2}A_{i}=0\\
\label{mag3}
& \partial_{j}^{2}B=0
\end{align}
\ees
A dot over any function of space-time will mean a partial derivative with respect to time from now onward. The Lagrangian enjoys the following gauge invariance:
\bes
\bea
&& \delta A=\partial_{t}\alpha_1 \;;\; \delta A_{i}=\partial_{i}\alpha_2 \;;\;\delta_{\alpha} B=0 \label{mgauge}\\
\mbox{where } && \, \, \p_i \p_t (\alpha_1 - \alpha_2) = 0 \label{mgauge_cond}
\eea
\ees
\subsection{Canonical analysis}
While trying to work out a Legendre transformation from this Lagrangian \eqref{Lmag}
we encounter a couple of primary constraints:
\bes \label{primary_constraints}
\bea
&& C^1 = \pi^A \approx 0 \label{c1}\\
&& C^2_i = \pi^A_i +\frac{1}{2} \p_i B \approx 0 \label{c2}.
\eea
\ees
Note that these two constraints Poisson commute among themselves. This calls for introduction of Lagrange multipliers and the Hamiltonian takes the form:
\bea \label{naivehamil}
H = \int d^d x \left[(\pi^B)^2 +  \pi^A  u_1+\left( \pi^A _i +\frac{1}{2} \p_i B\right) u^i_2  -\frac{1}{2} \p_i A \p_i B + \frac{1}{8} \left( F_{ij}\right)^2 \right]
\eea 
The primary constraints should be preserved in time and this enforces introduction of more constraints:
\bes \label{secondary_constraints}
\bea 
&& \{C^1, H\} = -\frac{1}{2} \p^2 B := C^3 \approx 0 \label{c3}\\
&& \{C^2_i, H\} = \frac{1}{2} \p_k F_{ki} + \p_i \pi^B := C^4_i \approx 0 \label{c4}
\eea
\ees
Continuing with the Dirac consistency algorithm in search for new constraints we Poisson commute the new constraints with the Hamiltonian \eqref{naivehamil} again, to have:  
\bes \label{lagmulsolve}
\bea 
&& \{C^3, H_n\} = -\p^2 \pi^B  \approx 0  \label{c3a}\\
&& \{C^4_i, H_n\} = \frac{1}{2} \left( \p^2 u^i_2 - \p_i\p^2 A\right) \approx 0 \label{c4a}
\eea
\ees
However \eqref{c3a} is not an independent constraint, as $\p \cdot C^4 = \p^2 \pi^B$, owing to the anti-symmetry of $f_{ij}$. On the other hand \eqref{c4a} is an equation for $u^i_2$ which can be solved as:
\bea \label{u2}
u^i_2 = \p_i A 
\eea
up to non-dynamical zero modes. Hence there are no more constraints. 

In order to classify the constraints into first and second class, we will smear them with appropriate vector or scalar test functions $\lambda$ and denote them as $\mathcal{C}_A[\lambda]$ for $A =1, \dots 4$.
For convenience, we make a redefinition for $\mathcal{C}_3[\lambda]$ in particular as:
\bea
\mathcal{C}_3[\lambda] = -\frac{1}{2} \int d^d x \, \lambda \p^2 B = \int d^d x \, \lambda \p_i \pi^A_i
\eea
where we have used the constraint $C^2_i$. Now, among the constraints the only non-vanishing bracket appears as:
\bea
\{\mathcal{C}_2[\lambda_2], \mathcal{C}_4[\lambda_4]\} = - \frac{1}{2} \int d^d x\,\lambda^i_2 \p^2 \lambda^i_4
\eea
making them second class while $\mathcal{C}_1, \mathcal{C}_3$ remain a first class set. This makes the physical phase space dimension = $2\times (2+d) - 2\times 2 - 2d = 0$, per space-time point. This makes it a topological field theory.

We would now see the gauge transformations generated by the first class constraints. In order to do that, let us rewrite the Hamiltonian \eqref{naivehamil} in terms of the solved $u^i_2$ \eqref{u2}:
\bea \label{finalhamil}
H = \int d^d x \left[(\pi^B)^2 +  u_1 \pi^A  + \p_i A \pi^A _i + \frac{1}{8} \left( F_{ij}\right)^2 \right]
\eea
An arbitrary gauge generator will be a linear combination of the first class constraints, ie:
\bea
G = \int d^d x \left(\lambda_1 \pi^A + \lambda_2 \p_i \pi^A_i \right)
\eea
But as we know from Dirac's algorithm, $\lambda_1$ and $\lambda_2$ can't both be independent gauge parameters, as there is one primary first class constraint. Their mutual dependence can be worked out by the off-shell condition \cite {Banerjee:1999hu}:
\bea \label{cond}
\delta_G \frac{d}{d t} q = \frac{d}{d t} \delta_G q
\eea 
where $q$ is any dynamical variable and $\delta_G$ is transformation generated by $G$ via:
$$\delta_G F(q,p) = \{F(q,p),G\}$$
for any phase space function $F$. For example:
\bea
\delta_G B = 0 ,\, \delta_G A = \lambda_1, \, \delta_G A_i = -\p_i \lambda_2
\eea
Implementing the above condition \eqref{cond} for the variables $B, A, A_i$ we obtain:
\bea
\p_i \p_t \lambda_2  = - \p_i \lambda_1.
\eea
This a cross-check that the gauge invariance mentioned in \eqref{mgauge}, \eqref{mgauge_cond} is reproduced correctly via canonical methods.
\subsection{Pre-symplectic analysis of symmetries}
On the space of solutions, which is plagued with gauge redundancy, we can introduce the presymplectic structure arising from the above Lagrangian:
\begin{align}
\label{msymp}
\Omega(\delta_1,\delta_2)=\frac{1}{2}\int d^d x\;[\;\delta_1 B.\partial_{t}\delta_2 B-\delta_1 A_i.\partial_{i}\delta_2 B- (1 \leftrightarrow 2)\;]
\end{align}
where $\delta_1, \delta_2$ are arbitrary variations, interpreted as tangent vector fields on the solution space are contracted with $\Omega$. With this structure, the space of solution is now a covariant phase space, an infinite dimensional (pre)-symplectic manifold. It is easy to see that the gauge transformation \eqref{mgauge} is a degenerate direction for $\Omega$ in a sense that:
$$ \Omega(\d_{\alpha}, \d) = 0$$
for field-independent gauge parameters $\alpha_1, \alpha_2$.
\subsubsection{M transformations}
Our next objective would be to see how the GCA generators act on the covariant phase space. For the sake of brevity we repackage the $M$ transformations appearing in \eqref{galconfinv2} with the constraints on the parameters \eqref{parameters} relevant for the magnetic sector:
\bes \label{transM}
\bea
&& \d_{\xi} B = \xi^k  \p_k B \\
&& \d_{\xi} A = \xi^k  \p_k A + A_k \dot{\xi}^k \\
&& \d_{\xi} A_i = \xi^k  \p_k A_i + B \dot{\xi}_i
\eea
\ees
For some spatially constant vector field $\xi^i (t)$, Laurent polynomial in time. 
It is easy to check that $\d_{\xi}$ as a tangent vector field on the covariant phase-space is locally Hamiltonian \footnote{For a symplectic manifold $(M,\omega)$, a vector field $X$ that preserves the symplectic 2-form $\Omega$ is said to be be generating symplectomorphism or simply Hamiltonian vector field. The condition is $ \pounds_X \Omega = 0$. Now $\Omega$, by definition being a closed 2-form, this implies $d(i_X \Omega) = 0$. If the first de-Rahm cohomology of trivial, then it guarantees a globally defined Hamiltonian function $i_X \Omega = dH_X$. Otherwise the existence of such Hamiltonian function would be a local statement. In the infinite dimensional covariant phase space context, this amounts to setting the criterion $\Omega (\delta_{\star}, \delta) = \delta Q_{\star}$, for the symplectomorphism $\delta_{\star}$ to be locally Hamiltonian.}
\bea
\Omega(\d_{\xi}, \d ) = \d \int d^d x \, \p_i B \left( \xi^i \p_t B - B \p_t \xi^i - \xi^k \p_k A_i \right)=:\d Q[\xi].
\eea
However, upon using the equations of motion \eqref{mag2}, \eqref{mag3} and eliminating total spatial divergence terms, it can be easily seen that the Hamilton function $Q[\xi] = 0$.

Therefore it may be tempting to view the transformations to be trivial gauge transformations, but as they don't vanish on-shell \eqref{meom}, they are gauge invariance of dynamical significance. As a special case, where $\dot{\xi}^i = 0$, ie for global spatial translations, the conserved total momentum of the system is zero, and remains so for all times. This observation, on the other hand, is consistent with the fact that the system is devoid of any local propagating degree of freedom, as deduced from the constraint analysis above.
\subsubsection{L transformations}
For transformations generated by $L$ \eqref{galconfinv2}, we would be a bit more adventurous enthused by the fact that even with restricted set of parameters \eqref{parameters}, only a finite number of them ($L^{(0, \pm 1)}$) generate symmetry. We are interested to see if the rest of generators generate canonical transformation on the phase space. Therefore we again relax the conditions \eqref{parameters} and bring in more generalities in the transformation rules:
\bes \label{transL}
\bea
&& \d_{f} B = f \dot{B}+ \dot{f} \left(  x^i \p_i + \Delta_1 \right)B -b'  \ddot{f} x^i A_i \\
&& \d_{f} A_i = f \dot{A}_i+ \dot{f} \left( x^j \p_j + \Delta_2 \right)A_i + \ddot{f} x_i \left( a'' B -a A \right) \\
&& \d_{f} A = f \dot{A}+ \dot{f} \left( x^i \p_i + \Delta_3 \right)A  -b \ddot{f} x^i A_i.
\eea
\ees
Here we have kept the conformal dimensions of the fields independent of each other \footnote{From the Lagrangian though it is evident that all the scaling dimensions should be identical for scale invariance. However, as the set of canonical transformations usually are a much bigger set than symmetry ones, we relax this.} and also used the transformation parameter as an arbitrary Laurent polynomial $f$ in time $t$.
These transformations \eqref{transL} generate Hamiltonian flow on the pre-symplectic phase space regardless of the form of $f$, in the sense that the form $ \Omega(\d_{f}, \d )$ is exact on the phase space, ie. integrable:
\bea
&&\Omega(\d_{f}, \d ) = \d Q[f] \non
\label{chargeL}
\eea
if and only if 
\bea \label{conditions}
b' = 0 = a , \, a'' = 1 ,\, \Delta_1 = \frac{ d -1}{2} = \Delta_2.  
\eea
However no restriction is put on the transformation properties of $A$, which is easily appreciated as $A$ does not enter the pre-symplectic structure explicitly. The spatial dimension $d$ is unrestricted though. The locally Hamilton function is given by:
\bea
 Q[f] &=& \frac{1}{2}\int d^d x \Bigg[ \frac{1}{2}f \left(\dot{B}^2+\frac{1}{2}f_{ij}f^{ij} \right)  + \frac{d +1}{4} \ddot{f} B^2 \non\\
&& +\dot{f} 
\left( \frac{d+1}{2} \left(A_i \p_i B  - B\dot{B}\right) +x^j \left( A^i \p_i \p_j B - B\p_j \dot{B} \right) \right)
\Bigg]
\eea

Now an explicit check gives us that $Q[f]$ can qualify as a conserved charge, ie $ \dot{Q}[f] = 0$ only if $d=3$ and $ \dddot{f} = 0$. This corroborates our earlier analysis of the symmetries at the level of equations of motion as well as the fact that out of infinite generators characterized by arbitrary function of time $f(t)$, only 3 independent ones  corresponding respectively to $L^{(-1)}, L^{(0)}, L^{(+1)}$, qualify as true symmetry of the theory \eqref{Lmag}. 

This is rather easily confirmed by directly acting \eqref{transL} on the Lagrangian \eqref{Lmag}. In addition to the conditions \eqref{conditions} and $d=3$, $\Delta_3 =1$ and $ b = -1$ has to be set to get off-shell:
\bea
\d_f L = \p_t \left[ f L +\frac{1}{4} \ddot{f} \int d^3 x B^2\right] + \frac{5}{4} \dddot{f} \int d^3 x B^2.
\eea 
This reestablishes the fact that only for $\dddot{f}=0$, the transformations are symmetries, in 3 spatial dimensions. Any such function can be expressed in the linearly independent basis of $f =1, t , t^2$. These correspond to 
time translation ($L^{(-1)}$ or $H$), dilatation ($L^{(0)}$ or $D$) and temporal part of Galilean special conformal transformation ($L^{(+1)}$ or $K$) respectively.
\subsubsection{Algebra of charges}
We have just observed that the $L$ and $M$ transformations of GCA produce (locally) Hamilton flows on the phase space. As an analogue of second part of Noether's theorem, we would like to verify whether the moment maps from GCA \eqref{gcafunc} to Hamilton functions \eqref{chargeL} are Lie-algebra homomorphisms. This is easily checked via
\bea \label{virasoro}
\Omega (\d_{f_1},\d_{f_2}) &=& \d_{f_2} Q[f_1] \non \\
&=& Q[f_3 = f_1 \dot{f}_2 - f_2 \dot{f}_1] + \frac{1}{2} (\dddot{f}_1 f_2-f_1 \dddot{f}_2) \int d^3 x B^2 .
\eea
This indicates that in case of $\dddot{f}_1 = 0=\dddot{f}_2$, the homomorphism exists, ie this is as expected for the f-GCA part of the $L$ generators, which are also symmetries of the theory. For arbitrary functions $f_1, f_2$ though, the last term above is a central one. If time $t$ is made to take value only on a bounded interval on real line and all the bosonic functions are taken to be single values functions of time, the above algebra \eqref{virasoro} becomes the Virasoro algebra with the state dependent central extension $\int B^2$.

On the other hand for the realization of the homomorphism from from the bracket $[L_{f}, M_{\xi}] = M_{\dot{f}\xi - f \dot{\xi}}$, we get trivially that:
\bea
\Omega (\d_{f} , \d_{\xi}) = \d_{\xi} Q[f] = 0
\eea
which is expected, as Hamilton functions corresponding to $M$ transformations have been found to be zero.
\section{Conclusions}
Let us briefly review what we have achieved in this paper. Firstly, we initiate the project of constructing an action starting from a set of equations of motion that describes physics on a degenerate manifold, focusing particularly to the case of Galilean field theories. For a free theory, we start from the most general set of second order differential equations for the fields, impose the Helmholtz conditions, keep only the terms which would give back the original equations of motion and satisfy the corresponding symmetries. If no solution consistent with all the three conditions are found, we add terms corresponding to an extra (scalar or vector) field to our set of most general second order differential equations and repeat the procedure. Using this procedure, we obtain the unique action for the magnetic sector of Galilean electrodynamics with the addition of an extra scalar field. The infinite dimensional group of global Galilean conformal symmetry enjoyed by the equations of motion gets broken, due to inclusion of this new scalar degree of freedom. However the symmetry algebra still remains an infinite dimensional (including conformal symmetry) subalgebra of the Galilean conformal algebra. We also show that it is not possible to obtain an action for the electric sector even if we add an extra scalar or an extra vector field and keep the system non-interacting. This observation clearly breaks the conventional wisdom of having electro-magnetic duality, which is otherwise observed in the relativistic case. 

Couple of comments are in order regarding our algorithm of finding an action for a Galilean theory. Firstly, we have kept conformal invariance to be a mandatory check. One could obviously get a more generic class of theories by relaxing the conformality condition. Secondly, we did not include half integer spins while looking for possible field extensions in the electric and the magnetic sector of fields. It would rather be an interesting study, especially in the electric sector, to see whether even with coupled spin-1/2 fermions one can get an action or not.

We used the action obtained in the magnetic limit to study the dynamical structure of the system in question, ie magnetic limit of Galilean electrodynamics. Dirac constraint analysis shows the non existence of propagating degrees of freedom in the system. Further we probed into those generators which are symmetries at the level of equations of motion, but did not survive after inclusion of an extra scalar field, at the level of action. In the covariant phase space framework, those generators still continue to be symmetries of the phase space and give rise to locally Hamiltonian flows, however only in 4 dimensional space-time. 

In our opinion, these questions regarding dynamics of Galilean field theories in general (conformal or otherwise) are important, because answers to these will strengthen a powerful conjecture (fueled by a large number of recent findings) made in \cite{Bagchi:2017yvj} that basically claims that every relativistic field theory in any dimension does have a Galilean subsector and that enjoys an enhanced, infinite amount of Galilean symmetries.  

There are several pieces of puzzle emerging from our study that are worth further investigations.  Firstly, the particular magnetic sector of Galilean electrodynamics we have constructed, is devoid of propagating degrees of freedom, which indicates a collapse of QED photon polarization states as we go to the Galilean limit. This is particularly hard to understand from the scale, spin highest weight representation that we have chosen to classify our dynamical fields. It would be useful to make an explicit map between this representation of Galilean algebra with the mass, spin representation of Poincare algebra as one takes the Galilean limit ($c \rightarrow \infty$). An analogous phenomena of collapse of perturbative closed string states to a single open string state have been discussed in \cite{Bagchi:2019cay}, when one takes a similar singular limit on worldsheet geometry. 

Talking about absence of local degrees of freedom, the role of global or boundary ones as in 3D gravity, would be an important investigation  \cite{Brown:1986nw}. Galilean theories described so far, have largely been for non interacting fields. The most obvious avenue to explore would be to extend the study to interacting fields. As it happens routinely for topological field theories, bringing in newer couplings is a way to generate local degrees freedom  \cite{Deser:1982vy}. However, an almost exhaustive set of possible interacting gauge theories allowed under Galilean symmetries are given in \cite{Bagchi:2017yvj}, at the level of equations of motion. Clearly a thorough investigation regarding classifying this large number of Galilean gauge theories through action formulation would be a worthwhile task. 

So far, the infinite dimensional Galilean conformal symmetries have only been studied classically. One of the most interesting feature of about classically conformal gauge theories is their quantum anomaly structure. Apart from a handful few, all such example are known to be anomalous (which are again integrable models), particularly in terms of the conformal Ward identities. Although a number of generic results have appeared in recent times regarding finite anomalies in finite Galilean symmetries \cite{Jensen:2014hqa}, \cite{Jain:2015jla}, a full fledged quantum calculation for an explicit model like ours is missing. It would therefore be interesting to check whether the infinite (or even a finite portion of the) conformal symmetries survive under quantization of the fields. In case the symmetries do not survive, quantum anomalies will arise and we will have to come up with a mechanism to cancel the anomalies.
 
Another interesting scenario is the construction of supersymmetric field theories which are invariant under infinite dimensional Galilean Conformal transformations both at classical level and at quantum level. If such invariant theories exist, then we wish to explore the breaking of supersymmetry in those theories. Some progress has already been made in this direction and will be reported elsewhere.
\section*{Acknowledgement}
It's a pleasure to thank Arjun Bagchi for numerous discussions and inputs, who is also instrumental in developing the ongoing project on investigations into Galilean gauge theories. Discussions with Aditya Mehra, Glenn Barnich, Stephane Detournay are gratefully acknowledged. Udit Narayan Chowdhury is thanked for collaborations in the initial part of the project. 

RB acknowledges support by DST (India) Inspire award, the Belgian Federal Science Policy Office (BELSPO), OPERA award from BITS Pilani. Hospitality provided by IIT Kanpur during various parts of this project is also acknowledged by RB.
\appendix
\section{Transformations of fields under Galilean Conformal Algebra}
In the case of action containing more than one scalar fields or vector field, the transformation 
under Galilean conformal algebra is generalized form of \eqref{galconfinv}.
The transformations of gauge fields $A$ (scalar field), $A_i$(vector field) and 
additional scalar field ($B$) and vector fields ($B_i$) under Galilean Conformal Algebra is ,
 
\begin{eqnarray}\nonumber
[L^{(n)},A(t,x)]&=& t^n [t\partial_t + (n+1)x^j \partial_j + (n+1)\Delta ] A(t,x)\\\nonumber
                    &-& b n(n+1) t^{n-1} x^i A_i(t,x) - e n(n+1) t^{n-1} x^i B_i(t,x) \\ \nonumber
[L^{(n)},B(t,x)]&=& t^n [t\partial_t + (n+1)x^j \partial_j + (n+1)\Delta] B(t,x)\\\nonumber
                    &-& b' n(n+1) t^{n-1} x^i A_i(t,x)\\\nonumber
[L^{(n)},A_i(t,x)] &=& t^n [ t\partial_t + (n+1)x^j \partial_j + (n+1)\Delta ] A_i(t,x)\\ \nonumber
                    &-& a n (n+1) t^{n-1} x_i A(t,x)-  a' n (n+1) t^{n-1} x_i B(t,x)\\\nonumber
[L^{(n)}, B_i(t,x)] &=& t^n [ t\partial_t + (n+1)x^j \partial_j + (n+1)\Delta ] B_i(t,x)\\ \nonumber
                    &-& e' n (n+1) t^{n-1} x_i A(t,x)\\\nonumber                    
[M_i^{(n)},A{(t,x)}] &=& -t^{(n+1)} \partial_i A(t,x) + b (n+1) t^n A_i(t,x) + e (n+1) t^n B_i(t,x) \\\nonumber
[M_i^{(n)},B{(t,x)}] &=& -t^{(n+1)} \partial_i B(t,x) + b' (n+1) t^n A_i(t,x) \\\nonumber
[M_i^{(n)},A_j{(t,x)}] &=& -t^{(n+1)} \partial_i A_j(t,x) + a (n+1) t^n \delta_{ij} A(t,x) + a' (n+1) t^n \delta_{ij} B(t,x) \\ \nonumber
[M_i^{(n)},B_j{(t,x)}] &=& -t^{(n+1)} \partial_i B_j(t,x) + e' (n+1) t^n \delta_{ij} A(t,x) \label{galconfinv1} \\
\end{eqnarray}
where $a,b,e,a',b',e'$ are arbitrary constants and $\Delta$ is dimension dependent.

Here,  if the action contains extra vector field $B_i$ (in addition to A and $A_i$) and not containing the scalar field B, the terms in the transformation containing
B will vanishes. Similarly, in the case of action containing A, $A_i$ and B, the terms containing $B_i$ in the transformation vanishes.

\section{Action Formulation for Carrollian limit} \label{carrollian}
Carrollian electrodynamics \cite{Duval:2014lpa,Bagchi:2016bcd,Bagchi:2019xfx,Basu:2018dub} is another non relativistic scaling of
space-time coordinates given by
\begin{eqnarray}
 x_i \rightarrow  x_i  , t \rightarrow \epsilon t ,   \epsilon \rightarrow 0 \label{carrollscaling}
\end{eqnarray}
Similar to the Galilean case, it also has two ways of scaling the gauge field components
\begin{eqnarray}
 \text{Magnetic limit} : A \rightarrow A ; A_i \rightarrow \epsilon A_i \\
\label{carrollmagnetic1}
 \text{Electric limit} : A \rightarrow \epsilon A ; A_i \rightarrow A_i \label{carrollelectric1}
\end{eqnarray}
The equations of motion for these limits are given by (here we keep the spatial dimension $d$ arbitrary)
\begin{eqnarray}
\mbox{Electric Limit: }~ && T_0 := \partial_j \partial_j A - \partial_j \partial_t A_j =
0\nonumber\\
&& T_i := - \partial_t \partial_i A + \partial_t \partial_t A_i = 0 \label{carrollelectric2}\\
\mbox{Magnetic Limit: }~ && T_0 = \partial_j \partial_t A_j = 0 \nonumber\\
&& T_i := \partial_t \partial_t A_i = 0 \label{carrollmagnetic2}
\end{eqnarray}
Under Carrollian conformal symmetries, the fields transform as {\cite{Basu:2018dub}}
\begin{eqnarray}\nonumber
\mbox{Translation} : \delta_p {A(x,t)} &=& p^j \partial_j A(x,t) ; \delta_p A_i(x,t)=p^j \partial_j A_i(t,x)\\\nonumber
\delta_p {B(x,t)} &=& p^j \partial_j B(x,t) ; \delta_p B_i(x,t)=p^j \partial_j B_i(t,x)\\\nonumber
\mbox{Rotation} :\delta_\omega A(x,t) &=& \omega^{ij}(x_i\partial_j - x_j \partial_i) A(x,t)\\\nonumber
\delta_\omega B(x,t) &=& \omega^{ij}(x_i\partial_j - x_j \partial_i) B(x,t)\\\nonumber
\delta_\omega A_l(x,t)&=& \omega^{ij}[(x_i \partial_j - x_j \partial_i) A_l + \delta_{l[i} A_{j]}] \\\nonumber
\delta_\omega B_l(x,t)&=& \omega^{ij}[(x_i \partial_j - x_j \partial_i) B_l + \delta_{l[i} B_{j]}] \\\nonumber
 \mbox{Dilatation}:\delta_{\Delta}A(x,t)&=& (t\partial_t + x^j \partial_j + \Delta)A(x,t)\\\nonumber
 \delta_{\Delta} B(x,t)&=& (t\partial_t + x^j \partial_j + \Delta)B(x,t)\\\nonumber
 \delta_\Delta a_i(x,t) &=& (t\partial_t + x^j \partial_j + \Delta) A_i(x,t)\\\nonumber
\delta_\Delta B_i(x,t) &=& (t\partial_t + x^j \partial_j + \Delta) B_i(x,t)\\\nonumber
\mbox{Special conformal}: \delta_{\kappa} A(x,t) &=& 2 k^i[(\Delta x_i + x_i t \partial_t + x_i x^j \partial_j -\frac{x^j x_j}{2}\partial_i) A(x,t)  \\\nonumber 
      &+& \kappa_1 t A_i + \kappa_1' t B_i] \\\nonumber 
 \delta_{\kappa} B(x,t) &=& 2 k^i[(\Delta x_i + x_i t \partial_t + x_i x^j \partial_j -\frac{x^j x_j}{2}\partial_i) B(x,t)  \\\nonumber 
      &+& \lambda_1 t A_i ] \\\nonumber 
 \delta_\kappa A_l(x,t) &=& 2 k^i (\Delta x_i + x_i t \partial_t + x_i x^j \partial_j - \frac{x^j x_j}{2}\partial_i) A_l \\ \nonumber &+& 2 k_l x^j A_j - 2 k^i x_l A_i + 2 \kappa_2 k_l t A  + 2 \kappa_2' k_l t B \\\nonumber 
 \delta_\kappa B_l(x,t) &=& 2 k^i (\Delta x_i + x_i t \partial_t + x_i x^j \partial_j - \frac{x^j x_j}{2}\partial_i) B_l \\ \nonumber &+& 2 k_l x^j B_j - 2 k^i x_l B_i + 2 \lambda_2 k_l t A \\\nonumber 
\end{eqnarray}
We will use the formalism developed in Section \ref{helmholtzaction} to verify the existence of
action and to construct it, if it exists.

The electric limit is simpler and we will discuss it first. Verifying the Helmholtz conditions for 
\eqref{carrollelectric2}, we can see that
\begin{eqnarray}
&& \frac{\partial T_0}{\partial (A_i)_{ab}} = -1 ~ ~ ~ ; ~ 
\mbox{ for } a=i, b=t \nonumber \\
&& \frac{\partial T_i}{\partial (A)_{ab}} = -1 ~ ~ ~; ~ 
\mbox{ for } a=i, b=t \nonumber 
\end{eqnarray}
The other Helmholtz conditions, and invariance under Carrollian conformal transformations (when
$\kappa_1 = 0$ and $\kappa_2 = 1, \Delta=1, d =3$) can be easily checked. The Lagrangian obtained is
\begin{eqnarray}
L = \int d^3 x \left(-\partial_j A \partial_j A + 2 \partial_j A \partial_t A_j 
- \partial_t A_j \partial_t A_j \right) \label{carrollelectriclag}
\end{eqnarray}
which is same as the action obtained in \cite{Basu:2018dub} (upto an overall negative sign) and
hence it is unique.

Let us consider the magnetic limit. The first Helmholtz condition is violated since
\begin{eqnarray}
&& \frac{\partial T_0}{\partial (A_i)_{ab}} = 1  ~ ~ ~; 
\mbox{ for } a=i, b=t \nonumber\\
&& \frac{\partial T_i}{\partial (A)_{ab}} = 0 ~ ~ ~; 
\mbox{ for all } a,b \nonumber
\end{eqnarray}
Let us supplement these equations with an additional scalar field. The most general equations linear in the fields will be same as those given in equations \eqref{eommostgeneral}. Consistency under Helmholtz conditions will again give equations \eqref{eomgeneral}. To obtain the magnetic limit once
$B$ is set to zero, some of the remaining arbitrary parameters have to be zero and we are left
with
\begin{eqnarray}
T_0 &=& a_5 \partial_j \partial_j B + a_4 \partial_t \partial_t B = 0\nonumber\\
T_i &=& b_4 \partial_t \partial_t A_i + b_5 \partial_i \partial_t B =0 \nonumber\\
T_{B} &=& c_4 \partial_j \partial_j B + c_5 \partial_t \partial_t B +
 b_5 \partial_t \partial_j A_j = 0 \label{eomcarollianmagnetic}
\end{eqnarray}
It can be checked that these equations will not be invariant under Carrollian conformal
transformations for any choice of these parameters. Hence no action formalism exists for the
magnetic limit even with the addition of any other scalar field.

Let us add a new vector field $B_i$. The most general equations, linear in the fields and
satisfying the Helmholtz conditions are given in equations \eqref{eomelectric2}. However, to get
electric limit $a_3$ should be non zero in the first equation of \eqref{eomelectric2} while it has to
be zero in the second equation of \eqref{eomelectric2}. Hence it is not possible to construct an
action for this case even with the addition of a vector field.

\end{document}